\documentclass[12pt,a4paper]{article}
\usepackage{amsmath,amsfonts,latexsym,amssymb,amsthm}
\usepackage{curves}

\begin{document}
  
\setlength{\unitlength}{0.4pt}
\linethickness{0.15mm}          

\theoremstyle{plain}
\newtheorem{thm}{Theorem}
\newtheorem{lem}{Lemma}
\newtheorem{cor}{Corollary}
\newtheorem{Q}{Question}

\theoremstyle{definition}
\newtheorem{case}{Case}

\renewcommand{\thesection}{\Roman{section}}
\renewcommand{\H}{\mathcal{H}}
\newcommand{\Tr}{\operatorname{Tr}} 
\newcommand{\defeq}{\overset{ \text{def} }{=}}
\newcommand{\qleq}{\overset{ ? }{\leqslant}}

\title{Monotonicity with volume of entropy and of mean entropy 
       for translationally invariant systems as consequences of strong 
       subadditivity}
\author{Amanda R. Kay\footnote{Electronic mail:
        ark102@york.ac.uk}\hspace{0.5em} 
        and Bernard S. Kay\footnote{Electronic mail: bsk2@york.ac.uk} \\
       \it{Department of Mathematics, University of York,} \\ 
       \it{York YO10~5DD, UK}}
\date{}
\maketitle
\begin{abstract} 
We consider some questions concerning the monotonicity properties of
entropy and mean entropy of states on translationally invariant systems
(classical lattice, quantum lattice and quantum continuous). By taking
the property of strong subadditivity, which for quantum systems was
proven rather late in the historical development, as one of four primary
axioms (the other three being simply positivity, subadditivity and
translational invariance) we are able to obtain results, some new, some
proved in a new way, which appear to complement in an interesting way
results proved around thirty years ago on limiting mean entropy and
related questions.  In particular, we prove that as the sizes of boxes
in $\mathbb{Z}^{\nu}$ or $\mathbb{R}^{\nu}$ increase in the sense of set
inclusion, (1)~their mean entropy decreases  monotonically  and
(2)~their entropy increases monotonically.  Our proof of (2) uses the
notion of \emph{m-point correlation entropies} which we introduce and
which generalize the notion of \emph{index of correlation} (see e.g.
R.~Horodecki, Phys.~Lett.~A 187 p145 1994).   We mention a number of
further results and questions concerning monotonicity of mean entropy
for more general shapes than boxes and for more general translationally
invariant (/homogeneous) lattices and spaces than  $\mathbb{Z}^{\nu}$ or
$\mathbb{R}^{\nu}$.
\end{abstract}


\pagebreak


\section{INTRODUCTION} \label{S:intro}

The mid 1960's saw the beginning of an intense period of research into
the mathematical properties of the entropy of translationally invariant
states on translationally invariant (infinite Euclidean) systems, both
classical and quantum.  Amongst the questions which were of interest at
that time was the question of the existence of what we shall call in
this paper \emph{limiting mean entropy}.  A simple variant of this
question (see Corollary~\ref{C:limiting} below) is the question whether
the mean entropy of a box tends to a definite limit as the lengths of
each of its sides tend to infinity.  Here, by the \emph{mean entropy} of
a (finite) box, we simply mean its entropy divided by its volume.  (The
reader should be warned that in the literature referred to here, no
particular phrase is attached to this concept, and the term `mean
entropy' is used instead to denote what we call here `limiting mean
entropy'.) This variant had been proven to be true both for classical
systems by Robinson and Ruelle  in~\cite{rr67} and for quantum systems
by Lanford and Robinson in~\cite{lr68}.  However, there were important
reasons for wanting to prove variants of this result which involved more
general shapes than boxes, such as the variant known as `(limiting) mean
entropy in the sense of van Hove'~\cite{rr67}. This had been proven in
the classical case in the Robinson-Ruelle paper~\cite{rr67} as a
consequence of a general property called \emph{strong subadditivity}
(SSA).  The Lanford-Robinson paper~\cite{lr68} put forward the
conjecture that SSA held also in the quantum case but, in the absence of
a proof of this, could not immediately establish limiting mean entropy
in the sense of van Hove.  (It was in fact first proven for quantum
systems by Araki and Lieb~\cite{al70}.)  In fact, six years were to pass
before SSA was finally established for quantum systems by Lieb and
Ruskai~\cite{lru73}.

Here we recall that, if $\rho_{123}$ is a state on a Hilbert space which
is given to us as a triple tensor product of three preferred Hilbert spaces,
$\H=\H_{1}\otimes\H_{2}\otimes\H_{3}$, and if $\rho_{2}$, $\rho_{12}$,
and $\rho_{23}$ denote its partial traces over $\H_{1}\otimes\H_{3}$,
$\H_{3}$, and $\H_{1}$ respectively, then the property of SSA can be 
written as
  \begin{equation} \label{Eq:ssain}
      S(\rho_{123})+S(\rho_{2})\leqslant S(\rho_{12})+S(\rho_{23})
  \end{equation}
where for any density matrix $\rho$, $S(\rho)$ denotes its (von Neumann) 
entropy $-\Tr(\rho\log\rho)$.

Since then, it has become increasingly clear that SSA is a key property
of states on composite quantum systems, and in particular of
translationally invariant states on translationally invariant quantum
systems~\cite{l75,r94}. However, we feel that our understanding of the
significance of SSA has remained incomplete, partly because of the
historical accident that its discovery and first use were very much
bound up with the specific technical problem of generalizing  results on
limiting mean entropy from the case of simple boxes -- where it was not
needed -- to the case of van Hove -- where it was useful.  In an attempt
to partially remedy this situation, we have considered a number of
questions relating to monotonicity properties of entropy and of mean
entropy of boxes, bearing in mind the SSA property from the outset.  We
have found that, while SSA might not be \emph{needed} to establish
limiting mean entropy for the case of boxes, it can, in fact,  be used
with profit to throw new light on this result. Namely, we shall show in
this paper that, for translationally invariant states on translationally
invariant quantum systems, SSA implies the stronger result that the
\emph{mean entropy} of boxes \emph{decreases} monotonically as the boxes
increase in size in the sense of set inclusion.    We shall also mention
a number of results and (as far as we are aware) open questions
concerning monotonicity of mean entropy, suggested by our approach,
which concern more general shapes than boxes and more general
translationally invariant (/homogeneous) lattices and spaces than the
usual infinite Euclidean lattices and spaces. Finally, we shall give a
new proof of the known result that SSA implies that the \emph{entropy}
of boxes \emph{increases} monotonically, again as the boxes increase in
size in the sense of set inclusion.

We now explain our basic setting and list our results in detail.  We
begin with the following discrete and continuous versions of the
standard definition of a translationally invariant quantum system (see
for example~\cite{lr68,r69}). In the case of a lattice,
$\mathbb{Z}^{\nu}$, we define a \emph{region} $\Lambda$ to be a 
non-empty finite subset. In the case of a continuum, $\mathbb{R}^{\nu}$,
we define a \emph{region} $\Lambda$ to be a measurable set with finite
(non-zero) volume.  In either case, there is an assignment of a
separable Hilbert space $\H_{\Lambda}$ to each region, satisfying, in
the continuum case, the additional condition that this assignment be the
same for any two regions which differ by a region of zero volume. 
Further, this assignment satisfies the compatibility condition that if
two regions $\Lambda_{1}$ and $\Lambda_{2}$ are disjoint, then
$\H_{\Lambda_{1}\cup\Lambda_{2}}=\H_{\Lambda_{1}}
\otimes\H_{\Lambda_{2}}$, where, in the continuum case, two regions are
said to be \emph{disjoint} if their intersection has zero volume. We
define a \emph{state} mathematically to consist of a family
$\{\rho_{\Lambda}\}$ of density operators (positive trace-class
operators with trace $1$) on the Hilbert spaces $\H_{\Lambda}$ which are
compatible in the sense that, for disjoint  $\Lambda_{1}$ and
$\Lambda_{2}$,  
 \begin{equation} \label{Eq:rho}
     \rho_{\Lambda_{1}}=\Tr_{\Lambda_{2}}(\rho_{\Lambda_{1}\cup\Lambda_{2}}) 
   \end{equation} 
where for any region $\Lambda$, $\Tr_{\Lambda}$ means the partial
trace over $\H_{\Lambda}$.

We remark that it is well known that classical lattice systems can be
regarded as special cases of quantum lattice systems, where the density
matrices representing the state are simultaneously diagonal, and so
any result for a quantum lattice system will also be true for a
classical lattice system. However, our results below are not applicable
to classical continuous systems since property (A) below fails
(see~\cite{rr67}) in this case. 

In this paper we shall mainly confine our interest to situations where
not only the quantum system, but also the \emph{state} is translationally
invariant.   This means that for all regions $\Lambda$ and all
translations $\tau$ from the relevant translation group
($\mathbb{Z}^{\nu}$ for lattice systems or $\mathbb{R}^{\nu}$ for
continuous systems) there exists a unitary operator $U(\tau,\Lambda)$
from $\H_{\Lambda}$ to $\H_{\tau(\Lambda)}$ such that
   \begin{equation} \label{Eq:traninv}
      \rho_{\tau(\Lambda)}=U(\tau,\Lambda)\rho_{\Lambda}U(\tau,\Lambda)^{-1}.
   \end{equation}    

Given any state on a translationally invariant quantum system, we define
the \emph{entropy of a region $\Lambda$} to be the von Neumann entropy
of $\rho_{\Lambda}$, i.e.\
   \begin{equation} \label{Eq:entropy}
       S(\Lambda)\defeq-\Tr(\rho_{\Lambda}\log\rho_{\Lambda}).
   \end{equation} 
The entropy of a region is known to satisfy many
properties~\cite{l75}. However, in the present paper we shall focus on: 
\bigskip

\noindent \emph{(A) Positivity.}
         \begin{equation*}
             S(\Lambda)\geqslant 0 \text{\qquad for all } \Lambda
         \end{equation*}

\noindent \emph{(B) Subadditivity (SA).}\\
\phantom{(B)} If $\Lambda_{1}$ and $\Lambda_{2}$ are disjoint, then
         \begin{equation*}
             S(\Lambda_{1}\cup\Lambda_{2})\leqslant
             S(\Lambda_{1})+S(\Lambda_{2})
         \end{equation*}

\noindent \emph{(C) Strong subadditivity (SSA).}
         \begin{equation*}
             S(\Lambda_{1}\cup\Lambda_{2}) 
            +S(\Lambda_{1}\cap\Lambda_{2})\leqslant
             S(\Lambda_{1})+S(\Lambda_{2}).
         \end{equation*}                        

\noindent  (A) follows  immediately from \eqref{Eq:entropy}. (C) follows
immediately from \eqref{Eq:ssain}, \eqref{Eq:rho}, and
\eqref{Eq:entropy}. (B) is just a special case of (C) but we prefer to
view it as a separate property.  Furthermore, if
our state is translationally invariant, it follows immediately  from
\eqref{Eq:traninv} and \eqref{Eq:entropy} that
\bigskip

\noindent \emph{(D) Translational invariance.}\\  
\phantom{(D)} For any element $\tau$ of the relevant translation group
         \begin{equation*} 
             S(\Lambda)=S(\tau(\Lambda)).
         \end{equation*} 

As we discussed above, Property (C) (or rather \eqref{Eq:ssain} from which
it is an easy consequence) has the status of a difficult
theorem~\cite{lru73}, but in spite of this, the game we wish to play
from now on is to regard (A), (B), (C) and (D) as axioms and to see what
one can easily prove about the class of functions $\Lambda\mapsto
S(\Lambda)$ from regions of $\mathbb{Z}^{\nu}$ or $\mathbb{R}^{\nu}$ to the
real numbers which obey these axioms.

We begin by defining the
\emph{mean entropy} $\bar{S}$ of a region $\Lambda$ by
   \begin{equation*}
         \bar{S}(\Lambda)\defeq\frac{S(\Lambda)}{|\Lambda|}
   \end{equation*}
where $|\Lambda|$ denotes, in the lattice case, the number of lattice points 
contained in $\Lambda$ and, in the continuum case, the volume of $\Lambda$. 

We also define the notion of \emph{box} regions, $\Lambda_{a}$,  
$a=(a_{1},\dots,a_{\nu})$, where $a_{1},\dots,a_{\nu}$ are positive
integers (in the lattice case) or positive real numbers (in the continuum
case) by
   \begin{equation*}
     \Lambda_{a}\defeq\lbrace x\in\mathbb{Z}^{\nu}\text{ or }\mathbb{R}^{\nu}:
      0<x_{i}\leqslant a_{i}\text{ for }i=1,\dots,\nu\rbrace
    \end{equation*}
These have $|\Lambda_{a}|=\prod_{i=1}^{\nu}a_{i}$. 
With these two definitions, we shall prove in Sections~\ref{S:thm1} 
and~\ref{S:correlate} that, both in the lattice and continuum cases, and
for arbitrary dimension $\nu$, Axioms (A), (B), (C) and (D) imply:

\begin{thm} \label{Th:meanent}
  \qquad $\Lambda_{a}\subset\Lambda_{b}\qquad\Rightarrow\qquad 
    \bar{S}(\Lambda_{a})\geqslant\bar{S}(\Lambda_{b})$
\end{thm}

\begin{thm} \label{Th:entropy}
  \qquad  $\Lambda_{a}\subset\Lambda_{b}\qquad\Rightarrow\qquad 
    S(\Lambda_{a})\leqslant S(\Lambda_{b})$
\end{thm} 

By Axiom (A) and the elementary result from real analysis that any
monotonic sequence which is bounded below has a limit, we immediately
have from Theorem~\ref{Th:meanent} the corollary:

\begin{cor} \label{C:limiting}
  Given any infinite sequence of boxes $\Lambda_{i}$, $i=1,2,\dots$, which 
  increase in size in the sense of set inclusion
     \begin{equation*}
        \lim_{i\rightarrow\infty}\bar{S}(\Lambda_{i})
     \end{equation*}
  exists.
\end{cor}  

\noindent The special case of this where every edge length of
$\Lambda_{i}$ tends to infinity as $i$ tends to infinity, is the result
of Lanford and Robinson~\cite{lr68}.

We have found a number of intriguing hints that it should be possible to
considerably generalize Theorem~\ref{Th:meanent} both to settings which
involve a class of shapes more general than boxes and to translationally
(and rotationally etc.) invariant systems more general than 
$\mathbb{Z}^{\nu}$ and $\mathbb{R}^{\nu}$. In Section~\ref{S:hexagon} we
outline a number of partial results in this direction and pose a number
of open questions.

Theorem~\ref{Th:entropy} is not an entirely new result.  Robinson and
Ruelle~\cite{rr67} proved such a monotonicity result, for classical
lattice systems, which was more general in that our boxes were replaced
by general regions. Also, in an article by Wehrl~\cite{w78},
Theorem~\ref{Th:entropy} is proven in the one-dimensional quantum case;
this can then easily be extended to higher dimensions as in our proof
below. However, we remark that Wehrl's proof \emph{both} relies on SSA
\emph{and} requires the existence (on the line) of limiting mean entropy
to have been established first. Instead, our proof proceeds directly
from Axioms (A), (B), (C) and (D) and involves the concept of
\emph{$m$-point correlation entropies} which we introduce in
Section~\ref{S:correlate} and which are related to the \emph{index of
correlation} (see~e.g. \cite{h94}) in somewhat the same way that
truncated correlation functions are related to full correlation
functions in quantum field theory and statistical mechanics 
\cite{ha96,p88}.


\section{PROOF OF THEOREM~\ref{Th:meanent}} \label{S:thm1}

We shall treat in turn the four cases of the one-dimensional lattice, the
$\nu$-dimensional lattice, the one-dimensional continuum, and the
$\nu$-dimensional continuum.  
 
\begin{case}[one-dimensional lattice] \label{case1}
In this case, a box region, $\Lambda_{(n)}$, is simply a set consisting of 
the first $n$ natural numbers for some natural number $n$.  Writing $S(n)$ 
instead of $S(\Lambda_{(n)})$ for ease of notation, it follows from Axioms 
(B) and (D) that
   \begin{equation} \label{Eq:SA2}
          S(q+r)\leqslant S(q)+S(r) 
   \end{equation}
and from Axioms (C) and (D) that
   \begin{equation} \label{Eq:SSA2}
          S(q+r+t)+S(r)\leqslant S(q+r)+S(r+t) 
   \end{equation}
where $q,r,t\in\mathbb{N}$.  

The statement of our theorem in this case
amounts to the statement that the mean entropy $S(n)/n$ is monotonically
decreasing.  We prove this by establishing the proposition
     \begin{equation} \label{Eq:prop}
         \frac{S(n)}{n}\geqslant\frac{S(n+1)}{n+1}
     \end{equation}
with the following simple inductive argument.  
First notice that a special case of \eqref{Eq:SA2} is
the statement that $S(2)\leqslant 2S(1)$. This establishes
Propostion~\eqref{Eq:prop} in the case $n=1$.
Next, on the assumption that Proposition~\eqref{Eq:prop} is true for $n=p$, 
we have, by~\eqref{Eq:SSA2} in the case $r=p$ and $q=t=1$ that
     \begin{align*}
          S(p+2)  &\leqslant S(p+1)+S(p+1)-S(p) \\
                  &\leqslant 2S(p+1)-\frac{p}{p+1}S(p+1) \\
                  &=\frac{p+2}{p+1}S(p+1)
     \end{align*}
which implies that~\eqref{Eq:prop} is true for $n=p+1$.  We conclude
that~\eqref{Eq:prop} is true and hence that Theorem~\ref{Th:meanent} 
is true in the case of a one-dimensional lattice.
\end{case}

\begin{case}[$\nu$-dimensional lattice] \label{case2}

With a similar change in notation to that used above, we now need to
prove
   \begin{equation}\label{Eq:case2}
        \frac{S(a_{1},\dots,a_{\nu})}{a_{1}\dots a_{\nu}}
        \geqslant
        \frac{S(b_{1},\dots,b_{\nu})}{b_{1}\dots b_{\nu}}
   \end{equation}
where $a_{i},b_{i}\in\mathbb{N}$ and $a_{i}\leqslant b_{i}$ for
$i=1,\dots,\nu$.    
We first notice that the function $S_{a_{2},\dots,a_{\nu}}(\cdot)\defeq
S(\cdot,a_{2},\dots,a_{\nu})$, from the natural numbers         
to $\mathbb{R}$, clearly satisfies~\eqref{Eq:SA2} 
and~\eqref{Eq:SSA2}. Thus, by Case~\ref{case1}, we have
   \begin{equation*}
        \frac{S(a_{1},a_{2},\dots,a_{\nu})}{a_{1}a_{2}\dots a_{\nu}}
        \geqslant
        \frac{S(b_{1},a_{2},\dots,a_{\nu})}{b_{1}a_{2}\dots a_{\nu}}
    \end{equation*}
We next notice that, in a similar way to above, the function
$S_{b_{1},a_{3},\dots,a_{\nu}}(\cdot)\defeq
S(b_{1},\cdot,a_{3},\dots,a_{\nu})$ also satisfies~\eqref{Eq:SA2} 
and~\eqref{Eq:SSA2}. Thus, by applying Case~\ref{case1} again, we have
    \begin{equation*}
        \frac{S(b_{1},a_{2},a_{3},\dots,a_{\nu})}
        {b_{1}a_{2}a_{3}\dots a_{\nu}}
        \geqslant
        \frac{S(b_{1},b_{2},a_{3},\dots,a_{\nu})}
        {b_{1}b_{2}a_{3}\dots a_{\nu}}
    \end{equation*}
One may clearly continue in this way, arriving at~\eqref{Eq:case2}
after a total of $\nu$ such steps.
\end{case}

\begin{case}[one-dimensional continuum] \label{case3}
In this case, a box region, $\Lambda_{(x)}$, is simply a real interval
$(0,x]$. Writing $S(x)$ instead of $S(\Lambda_{(x)})$ we now need to
prove
     \begin{equation} \label{Eq:case3}
           \frac{S(y)}{y}\geqslant\frac{S(x)}{x}
     \end{equation}  
for $y\leqslant x$.  

We first argue that~\eqref{Eq:case3} holds on the rationals.  For any
two rationals $x$ and $y$, let $c$ be their common denominator and
define the function $S_c(\cdot)$, taking its argument from the natural
numbers, by $S_c(n)\defeq S(n/c)$. This function
satisfies~\eqref{Eq:SA2} and~\eqref{Eq:SSA2} of Case~\ref{case1} and
thus $S_{c}(n)/n$ and hence $S(n/c)/(n/c)$ are monotonically decreasing
by the argument given there, thus establishing~\eqref{Eq:case3} for
these $x$ and $y$.  To extend~\eqref{Eq:case3} to the reals, it then
clearly suffices to prove that $S(x)$ is continuous.  This follows
immediately from the following lemmas and Axiom (A).

\begin{lem}[Lieb] \label{L:lieb} 
    $S(x)$ is weakly concave i.e.\ for positive real numbers $x$ and $y$, 
    $S((x+y)/2)\geqslant S(x)/2 +S(y)/2$.  
\end{lem} 

\begin{lem} \label{L:cont}
    A function which is weakly concave and bounded below is necessarily 
    continuous. 
\end{lem}           

\noindent To prove Lemma~\ref{L:lieb}, first note that if $x=y$ the
statement is  trivially true. Otherwise, assume without loss of
generality that $y< x$.  The result then  follows from~\eqref{Eq:SSA2}
in the case established above where $q,r$ and $t$ are real, by
identifying $q=t=(x-y)/2$ and $r=y$. We remark that this is essentially
the same as an argument given in~\cite{w78},  where it is attributed to
E. Lieb.  Lemma~\ref{L:cont} (or rather the alternative statement with
``convex'' substituted for ``concave'' and ``bounded above'' substituted 
for ``bounded below'') is proved in~\cite{ps72}. We remark that this is
the only place where we use Axiom (A). In particular, Axiom (A) is
unnecessary for Cases~\ref{case1} and~\ref{case2}.
\end{case}

\begin{case}[$\nu$-dimensional continuum] \label{case4}
This case can be established from Case~\ref{case3} by an argument similar to 
that used above to go from Case~\ref{case1} to Case~\ref{case2}.
\end{case}

This completes the proof of Theorem~\ref{Th:meanent}. We remark that it
can be helpful to visualize the steps in the above proof using a
geometrical picture in which lattice points are identified with 
$\nu$-dimensional continuum cubes of side $1$.  In detail,
one identifies the particular lattice point $(1,\dots,1)$ with the 
particular continuum cube $\Lambda_{(1,\dots,1)}$ and extends this
identification by identifying the general lattice point
$(a_{1},\dots,a_{\nu})$, $a_{1},\dots,a_{\nu}\in\mathbb{Z}$, with the
result of translating the cube $\Lambda_{(1,\dots,1)}$ by the vector
$(a_{1}-1,\dots,a_{\nu}-1)$.  We also remark that, in the continuum
case, Theorem~\ref{Th:meanent} can trivially be extended from the case
of nested box  regions to  nested parallelepiped regions with parallel
faces (by simply ``squashing'' the boxes in the theorem).


\section{REMARKS ABOUT POSSIBLE GENERALIZATIONS OF THEOREM~\ref{Th:meanent}} 
\label{S:hexagon}

We now discuss two different directions in which one can attempt to
generalize Theorem~\ref{Th:meanent}.

Firstly, one can ask whether Theorem~\ref{Th:meanent} generalizes to
more general shapes than boxes (or parallelepipeds).  Indeed, one can ask 
the very general question 
 \begin{Q} \label{Q:gen} 
   Is mean entropy monotonically decreasing on any sequence of
   regions in $\mathbb{Z}^{\nu}$ or $\mathbb{R}^{\nu}$ which 
   increase in size in the sense of set inclusion? 
 \end{Q}
\noindent 
In other words, is the mean entropy of any region in the system less
than or equal to the mean entropy  of any subregion of that region?  We
remark that this question is more likely to have a positive answer if we
extend the translation group of Section~\ref{S:intro}
to the
appropriate full symmetry group of $\mathbb{Z}^\nu$ or $\mathbb{R}^\nu$,
i.e.\ if we also include rotations and reflections. From now
on we shall assume this extension to be made. We have been unable to
answer this question in anything like its full generality, but we have
found no negative answers and some partial positive answers in the case
of a  few specific simple shapes which go beyond the box-shapes (and
parallelepiped shapes -- cf. the second remark at the end of
Section~\ref{S:thm1}) of Theorem~\ref{Th:meanent}.   For example, in
$\mathbb{Z}^{2}$ we can prove inequalities such as
 \begin{equation} \label{Eq:boxes}
   \frac{S(  
         \begin{picture}(20,20)(0,0)
            \curve(0,0, 20,0)
            \curve(0,0, 0,20)
            \curve(0,10, 20,10)
            \curve(10,0, 10,20)
            \curve(20,0, 20,10)
            \curve(0,20, 10,20) 
         \end{picture}
                                    )}{3}
   \leqslant
   \frac{S(
         \begin{picture}(20,10)(0,0)
            \curve(0,0, 20,0)
            \curve(0,10, 20,10)
            \curve(0,0, 0,10)
            \curve(10,0, 10,10)
            \curve(20,0, 20,10)
         \end{picture}
                                    )}{2}        
 \end{equation}
where we are now using an obvious notation suggested by the first remark 
at the end of Section~\ref{S:thm1}.

Equation \eqref{Eq:boxes} may easily be proven from the special cases
 \begin{align*}
   S(\begin{picture}(20,10)(0,0)
          \curve(0,0, 20,0)
          \curve(0,10, 20,10)                                          
          \curve(0,0, 0,10)
          \curve(10,0, 10,10)
          \curve(20,0, 20,10)
     \end{picture})
   &\leqslant
   S(\begin{picture}(10,10)(0,0)
          \curve(0,0, 10,0)
          \curve(0,10, 10,10)                                          
          \curve(0,0, 0,10)
          \curve(10,0, 10,10)
     \end{picture})
   +
   S(\begin{picture}(10,10)(0,0)
          \curve(0,0, 10,0)
          \curve(0,10, 10,10)                                          
          \curve(0,0, 0,10)
          \curve(10,0, 10,10)
     \end{picture}) \\
   S(\begin{picture}(20,20)(0,0)
            \curve(0,0, 20,0)
            \curve(0,0, 0,20)
            \curve(0,10, 20,10)
            \curve(10,0, 10,20)
            \curve(20,0, 20,10)
            \curve(0,20, 10,20) 
      \end{picture})
   +
   S(\begin{picture}(10,10)(0,0)
            \curve(0,0, 10,0)
            \curve(0,10, 10,10)                                          
            \curve(0,0, 0,10)
            \curve(10,0, 10,10)
     \end{picture})
   &\leqslant
   S(\begin{picture}(20,10)(0,0)
          \curve(0,0, 20,0)
          \curve(0,10, 20,10)                                          
          \curve(0,0, 0,10)
          \curve(10,0, 10,10)
          \curve(20,0, 20,10)
     \end{picture})
   +
   S(\begin{picture}(20,10)(0,0)
          \curve(0,0, 20,0)
          \curve(0,10, 20,10)                                          
          \curve(0,0, 0,10)
          \curve(10,0, 10,10)
          \curve(20,0, 20,10)
     \end{picture})
 \end{align*}    
of subadditivity and strong subadditivity in an entirely analogous way
to the way we established Case~\ref{case1} of Theorem~\ref{Th:meanent} 
from equations~\eqref{Eq:SA2} and~\eqref{Eq:SSA2} in
Section~\ref{S:thm1} in the case that $q=r=t=1$.  However we have been
unable, for example, to prove or disprove either of the candidate
inequalities
\begin{align}
\frac{
S(\begin{picture}(30,20)(0,0)
           \curve(0,0, 30,0)
           \curve(0,10, 30,10)
           \curve(0,20, 10,20)
           \curve(0,0, 0,20)
           \curve(10,0, 10,20)
           \curve(20,0, 20,10)
           \curve(30,0, 30,10)
  \end{picture})                    }{4}
&\qleq
\frac{
S(\begin{picture}(20,20)(0,0)
            \curve(0,0, 20,0)
            \curve(0,0, 0,20)
            \curve(0,10, 20,10)
            \curve(10,0, 10,20)
            \curve(20,0, 20,10)
            \curve(0,20, 10,20) 
      \end{picture})                 }{3}  \label{Eq:lshape} \\
\frac{
S(\begin{picture}(30,20)(0,0)
           \curve(0,0, 30,0)
           \curve(0,10, 30,10)
           \curve(0,20, 10,20)
           \curve(0,0, 0,20)
           \curve(10,0, 10,20)
           \curve(20,0, 20,10)
           \curve(30,0, 30,10)
  \end{picture})                    }{4}
&\qleq
\frac{
S(\begin{picture}(30,10)(0,0)
           \curve(0,0, 30,0)
           \curve(0,10, 30,10)
           \curve(0,0, 0,10)
           \curve(10,0, 10,10)
           \curve(20,0, 20,10)
           \curve(30,0, 30,10)
  \end{picture})                    }{3}  \label{Eq:plank}
\end{align}
(but see after Corollary~\ref{C:av} below for a partial answer to these
questions).

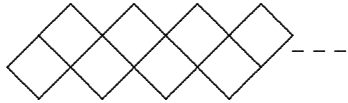
\begin{figure}[h!]
\begin{center}
\begin{picture}(320,90)(0,0)
  \curve(0,30, 60,90)
  \curve(30,0, 120,90)
  \curve(90,0, 180,90)
  \curve(150,0, 240,90)
  \curve(210,0, 270,60)
  \curve(30,0, 0,30)
  \curve(90,0, 30,60)
  \curve(150,0, 60,90)  
  \curve(210,0, 120,90)
  \curve(240,30, 180,90)
  \curve(270,60, 240,90)
  \curve(270,45, 280,45)
  \curve(290,45, 300,45)
  \curve(310,45, 320,45)
\end{picture}
\end{center}
\caption{Chain of figures}\label{Fig}
\end{figure}

In fact, many of the cases where we have been able to answer 
Question~\ref{Q:gen} positively turn out to refer to consecutive figures
in a one-dimensional ``chain'' of figures.  For example the
case~\eqref{Eq:boxes} illustrated above clearly easily extends to a more
general inequality which refers to an arbitrary pair of successive
figures in the chain shown in Figure~\ref{Fig}.

An interesting special case of Question~\ref{Q:gen} is 
 \begin{Q} \label{Q:spec}
    Is mean entropy monotonically decreasing on any sequence 
    of \emph{similar} regions in $\mathbb{Z}^{\nu}$ or $\mathbb{R}^{\nu}$ 
    which increase in size in the sense of set inclusion? 
 \end{Q}   

Of course, we know from Theorem~\ref{Th:meanent} that we can answer 
Question~\ref{Q:spec} positively  for the case of similar boxes and
parallelepipeds.    But consideration of more general shapes forces us
to leave the realm of one-dimensional chains and, for this reason, we
have found it  difficult to find other shapes for which we can prove
anything.   In fact, we have not even been able to answer
Question~\ref{Q:spec} in the case of discs in $\mathbb{R}^{2}$ with
increasing radii. However, we \emph{have} been able to answer
Question~\ref{Q:spec} positively in the case of two regular hexagons in
the plane (i.e.\ $\mathbb{R}^2$) with diameters in the ratio two-to-one.

\begin{figure}[!h] 
 \begin{center}
  \begin{picture}(160,140)(0,0) 
        \curve(40,138, 120,138)
        \curve(20,103.5, 140,103.5)
        \curve(0,69, 160,69)
        \curve(20,34.5, 140,34.5)
        \curve(40,0, 120,0)
        \curve(0,69, 40,138)
        \curve(20,34.5, 80,138)
        \curve(40,0, 120,138)
        \curve(80,0, 140,103.5)         
        \curve(120,0, 160,69)
        \curve(0,69, 40,0)
        \curve(20,103.5, 80,0)
        \curve(40,138, 120,0)
        \curve(80,138, 140,34.5)
        \curve(120,138, 160,69)
  \end{picture}
  \caption{Hexagon figure} \label{fighex}
 \end{center}
\end{figure}
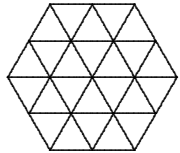

To treat this situation we consider Figure~\ref{fighex}.
Denoting the smaller hexagon made of 6 small triangles by $H$ and the 
diamond region made of two small triangles by $D$, we
begin by noting that the mean entropy of $H$ is less than or equal to
the mean entropy of $D$. This follows immediately once one notices  that
$H$ can be viewed as the disjoint union of three copies of $D$, since by
applying Axiom (B) (twice) we have $S(H)\leqslant S(D)+S(D)+S(D)$ which
implies that
   \begin{equation} \label{Eq:hexdiam}
       \frac{S(H)}{6}\leqslant \frac{S(D)}{2}
   \end{equation}

Next, imagine that the vertices of the central small hexagon  in
Figure~\ref{fighex} are numbered (say) clockwise, starting at some
particular vertex, from 1 to 6 and regard the large hexagon as the union
of 6 copies of $H$, which we shall call $H_1,\dots,H_6$, centred
respectively at each of these 6 vertices. Also define the sequence of
figures $F_1=H_1$, $F_2=F_1\cup H_2$, $F_3=F_2\cup H_3$, etc.\ so that
$F_6$ is our large hexagon.  We may then argue that each of these
figures $F_n$, taken successively, has a mean entropy less than or equal
to that of $H$. The first step in this argument proceeds by first
noticing that $F_2$ consists of the union of two copies of $H$ whose
intersection is a copy of $D$ and hence by Axioms (C) and (D) that 
$S(F_2)+S(D)\leqslant S(H)+S(H)$.  This is easily combined with the 
inequality~\eqref{Eq:hexdiam} to conclude that $S(F_2)/10\leqslant
S(H)/6$.  The subsequent steps proceed along similar lines, each using
the result of the previous step, along with inequality \eqref{Eq:hexdiam} 
and the facts that (a) $F_i$ consists of the union of the figure
$F_{i-1}$ and a copy of the figure $H$ (b) the intersection of $F_{i-1}$
and the same $H$ is a  copy of $D$. After the fourth step we have the
result that  $S(F_{5})/22\leqslant S(H)/6$. For the final step we note
that $F_6$ is the union of the figure $F_5$ and a  copy of the figure
$H$, but this time the intersection of these figures is a new figure
$G$ (composed of 4 small triangles). To derive the final result that
$S(F_{6})/24\leqslant S(H)/6$ we now need, instead of
\eqref{Eq:hexdiam}, the result that  $S(H)/6\leqslant S(G)/4$.
This can easily be shown by  using Axioms (C) and (D) to establish that
$S(H)+S(D)\leqslant 2S(G)$ and combining this with~\eqref{Eq:hexdiam}. 

Besides the above specific examples, we can also prove (say for a
lattice $\mathbb{Z}^{\nu}$, and continuing to interpret lattice points
as cubes and to refer to collections of cubes as `figures') the general
result: 

\begin{thm}  \label{Th:av.entropy}
    The mean entropy of a figure $\mathcal{F}(n)$ composed of $n$ cubes
    ($n\geqslant 2$) is less than or equal to the average of the mean 
    entropies of all the (connected or disconnected) figures contained 
    in $\mathcal{F}(n)$ which are composed of $n-1$ cubes. 
\end{thm} 

We remark that Theorem~\ref{Th:av.entropy} and Corollary~\ref{C:av}
below  actually only assume Axioms (B) and (C).  In particular, the
symmetry-invariance axiom (D) is not required in any form.

\begin{proof}[Proof of Theorem~\ref{Th:av.entropy}] 
First we introduce some new notation.
Labelling the cubes of $\mathcal{F}(n)$ by the integers $1,\dots,n$
we let $\mathcal{F}(n;i,j,\dots)$ denote the
figure that is formed from the figure $\mathcal{F}(n)$ by taking away
its $i$th, $j$th, $\dots$ cubes. Then the statement of  
Theorem~\ref{Th:av.entropy} amounts to
  \begin{equation} \label{Eq:av.entropy}
       \frac{S(\mathcal{F}(n))}{n}\leqslant\frac{1}{n}\sum_{j}
       \frac{S(\mathcal{F}(n;j))}{n-1}
  \end{equation}

We prove this inequality by induction on $n$.
First,~\eqref{Eq:av.entropy}  is true for all figures $\mathcal{F}$ with
$n=2$ by Axiom (B).  Next, we assume that~\eqref{Eq:av.entropy} is true
for all figures $\mathcal{F}$ with $n=p$ cubes. Taking any figure
$\mathcal{F}(p+1)$, we note  that $\mathcal{F}(p+1;i)$ consists of just
$p$ cubes, so by our assumption
  \begin{equation} \label{Eq:av.entropy2}
       \frac{S(\mathcal{F}(p+1;i))}{p}\leqslant\frac{1}{p}\sum_{j\neq i}
       \frac{S(\mathcal{F}(p+1;i,j))}{p-1}
  \end{equation}
Also, for $j\neq i$, Axiom (C) implies that
  \begin{equation} \label{Eq:SSA3}
      S(\mathcal{F}(p+1))\leqslant S(\mathcal{F}(p+1;i))+S(\mathcal{F}(p+1;j))
       -S(\mathcal{F}(p+1;i,j))
  \end{equation}
Summing~\eqref{Eq:SSA3} for $j=1,\dots,p+1$, with $j\neq i$, leads to
  \begin{multline*}
       pS(\mathcal{F}(p+1)) \leqslant  pS(\mathcal{F}(p+1;i)) \\ 
       +\sum_{j\neq i}S(\mathcal{F}(p+1;j)) 
       -\sum_{j\neq i}S(\mathcal{F}(p+1;i,j))
  \end{multline*}
Combining this with~\eqref{Eq:av.entropy2} we have
  \begin{align*}
       pS(\mathcal{F}(p+1))
             & \leqslant pS(\mathcal{F}(p+1;i))+\sum_{j\neq i}
                                   S(\mathcal{F}(p+1;j)) \\
             & \hspace{5cm}  -(p-1)S(\mathcal{F}(p+1;i)) \\
             &=\sum_{j}S(\mathcal{F}(p+1;j))
  \end{align*}
Dividing this last equation by $p(p+1)$ shows that~\eqref{Eq:av.entropy} is 
true for $n=p+1$. 
\end{proof}
 
This theorem also leads to the natural corollary:

\begin{cor} \label{C:av}
    \begin{equation*}
       \frac{S(\mathcal{F}(n))}{n}\leqslant
       \frac{\max_{j}S(\mathcal{F}(n;j))}{n-1}
    \end{equation*}  
\end{cor}

Thus the mean entropy of a figure on a lattice is less than or equal to 
the mean entropy of at least one of its subfigures composed of one
less cube. Returning to an example discussed above, we see that
this remark implies that the  mean entropy of the figure 
\begin{picture}(30,20)(0,0)
           \curve(0,0, 30,0)
           \curve(0,10, 30,10)
           \curve(0,20, 10,20)
           \curve(0,0, 0,20)
           \curve(10,0, 10,20)
           \curve(20,0, 20,10)
           \curve(30,0, 30,10)
\end{picture}
is less than or equal to the mean entropy of one of its four
subfigures each composed of 3 cubes. In fact, we have been able to prove,
by an alternative route, the stronger result that its mean entropy is
less than or equal to the mean entropy of one of the two
\emph{connected} subfigures composed of 3 cubes, i.e.\ that one of the two
inequalities~\eqref{Eq:lshape} and~\eqref{Eq:plank} is actually true, but
we can't say which one. This is done by first noting that by Axiom (C):
\begin{equation*}
   S(\begin{picture}(30,20)(0,0)
           \curve(0,0, 30,0)
           \curve(0,10, 30,10)
           \curve(0,20, 10,20)
           \curve(0,0, 0,20)
           \curve(10,0, 10,20)
           \curve(20,0, 20,10)
           \curve(30,0, 30,10)
     \end{picture})
  +
   S(\begin{picture}(20,10)(0,0)
            \curve(0,0, 20,0)
            \curve(0,10, 20,10)
            \curve(0,0, 0,10)
            \curve(10,0, 10,10)
            \curve(20,0, 20,10)
     \end{picture}) 
  \leqslant
   S(\begin{picture}(30,10)(0,0)
           \curve(0,0, 30,0)
           \curve(0,10, 30,10)
           \curve(0,0, 0,10)
           \curve(10,0, 10,10)
           \curve(20,0, 20,10)
           \curve(30,0, 30,10)
     \end{picture})
  +
   S(\begin{picture}(20,20)(0,0)
            \curve(0,0, 20,0)
            \curve(0,0, 0,20)
            \curve(0,10, 20,10)
            \curve(10,0, 10,20)
            \curve(20,0, 20,10)
            \curve(0,20, 10,20) 
     \end{picture})   
\end{equation*}
Combining this with~\eqref{Eq:boxes} we have:
\begin{equation} \label{Eq:three}
   S(\begin{picture}(30,20)(0,0)
           \curve(0,0, 30,0)
           \curve(0,10, 30,10)
           \curve(0,20, 10,20)
           \curve(0,0, 0,20)
           \curve(10,0, 10,20)
           \curve(20,0, 20,10)
           \curve(30,0, 30,10)
     \end{picture})
  \leqslant
   S(\begin{picture}(30,10)(0,0)
           \curve(0,0, 30,0)
           \curve(0,10, 30,10)
           \curve(0,0, 0,10)
           \curve(10,0, 10,10)
           \curve(20,0, 20,10)
           \curve(30,0, 30,10)
     \end{picture})
  +\frac{1}{3}
   S(\begin{picture}(20,20)(0,0)
            \curve(0,0, 20,0)
            \curve(0,0, 0,20)
            \curve(0,10, 20,10)
            \curve(10,0, 10,20)
            \curve(20,0, 20,10)
            \curve(0,20, 10,20) 
     \end{picture})  
\end{equation}
But, we must have \emph{either} 
$  S(\begin{picture}(30,10)(0,0)
           \curve(0,0, 30,0)
           \curve(0,10, 30,10)
           \curve(0,0, 0,10)
           \curve(10,0, 10,10)
           \curve(20,0, 20,10)
           \curve(30,0, 30,10)
     \end{picture})
 \leqslant 
  S(\begin{picture}(20,20)(0,0)
            \curve(0,0, 20,0)
            \curve(0,0, 0,20)
            \curve(0,10, 20,10)
            \curve(10,0, 10,20)
            \curve(20,0, 20,10)
            \curve(0,20, 10,20) 
     \end{picture})  $  
\emph{or}
$ S(\begin{picture}(20,20)(0,0)
            \curve(0,0, 20,0)
            \curve(0,0, 0,20)
            \curve(0,10, 20,10)
            \curve(10,0, 10,20)
            \curve(20,0, 20,10)
            \curve(0,20, 10,20) 
     \end{picture})
 \leqslant
  S(\begin{picture}(30,10)(0,0)
           \curve(0,0, 30,0)
           \curve(0,10, 30,10)
           \curve(0,0, 0,10)
           \curve(10,0, 10,10)
           \curve(20,0, 20,10)
           \curve(30,0, 30,10)
     \end{picture})  $. 
Thus, we conclude from~\eqref{Eq:three} that one of the  
inequalities~\eqref{Eq:lshape} and~\eqref{Eq:plank} must be true.

A second direction in which one can attempt to generalize
Theorem~\ref{Th:meanent} is suggested by the fact that the basic setting
of Section~\ref{S:intro} clearly generalizes to more general lattices
than $\mathbb{Z}^\nu$ and to more general homogeneous spaces than
$\mathbb{R}^{\nu}$ such as discs in one-dimension and spheres and tori
in higher dimensions. One can thus ask to what extent
Theorem~\ref{Th:meanent} generalizes to such settings, where Axiom (D)
is now replaced by invariance under the relevant symmetry group.   As
far as more general lattices are concerned, we remark that the hexagon
example discussed above could be regarded as an example concerning a
triangular lattice.  For the case of the one-dimensional circle and
higher dimensional tori, it is easy to see that the obvious analogue of
Theorem~\ref{Th:meanent} still holds. For example, on both a
one-dimensional ``lattice unit-circle'' (where the allowed angles are
$2m\pi/n$, $m=1,\dots,n$) and a ``continuum unit-circle'', one easily
shows by a close analogue to the arguments in Cases~\ref{case1}
and~\ref{case3} of Section~\ref{S:thm1} that 
   \begin{equation*}
     \frac{S(\theta_{1})}{\theta_{1}}\geqslant
     \frac{S(\theta_{2})}{\theta_{2}}
     \text{\qquad for \qquad}
     \theta_{1}\leqslant\theta_{2}
  \end{equation*}
It is natural to ask the following question (and the obvious
counterparts to this question in higher dimensions) concerning a
possible generalization of this result, in the continuum case, to the
2-sphere:  

\begin{Q}
Does the mean entropy of a disc drawn on the surface of a sphere
decrease monotonically as the solid angle subtended at the centre
increases?
\end{Q}

\noindent But, just as for discs in $\mathbb{R}^{2}$, we have been unable to
answer this question.
       

\section{PROOF OF THEOREM~\ref{Th:entropy}} \label{S:correlate}

We shall find it useful to begin by introducing, in the case of a
one-dimensional lattice, the notion of the  \emph{m-point correlation
entropies} of a translationally invariant state.
  
To motivate this definition, we first recall the notion of the
\emph{index of correlation} (see for example~\cite{h94} where it is
discussed in an
abstract setting concerning states on tensor products of Hilbert spaces).
In the case of a one-dimensional quantum lattice system we can
interpret this as the difference between the entropy of the union of $n$
consecutive lattice points (or, in our alternative interpretation, cubes) 
and the sum of their individual entropies:
  \begin{equation} \label{Eq:index}
       I_{n}\defeq n S(1)-S(n)
  \end{equation}
By using Axiom (B) $n-1$ times, it is easy to show that $I_{n}$ is 
positive.

Our new notion of \emph{m-point correlation entropies}  may be
regarded as designed so as to provide a new way of writing the index of
correlation $I_{n}$  as a sum of positive terms, each of which
concerns $m\leqslant n$ lattice points.  Namely, we define the
\emph{m-point correlation entropies} by
  \begin{equation} \label{Eq:corr.entropy}
       S^{c}_{m} \defeq  \begin{cases} 
                            2S(1)-S(2) & m=2  \\
                            2S(m-1)-S(m-2)-S(m) & m\geqslant 3 
                         \end{cases}
  \end{equation}
Note that $S^{c}_{m}$ is positive by Axiom (B) for $m=2$ and by Axiom
(C) for $m\geqslant3$. An easy calculation then shows that 
  \begin{equation} \label{Eq:index2}
       I_{n}=\sum_{m=2}^{n}(n+1-m)S^{c}_{m}
  \end{equation}
By~\eqref{Eq:index} and~\eqref{Eq:index2}, we can write the entropy of $n$ 
consecutive lattice points as 
  \begin{equation} \label{Eq:entropysum}
       S(n)=nS(1)-\sum_{m=2}^{n}(n+1-m)S^{c}_{m}
  \end{equation}
We note that by adding an extra lattice point onto a region of $n$
consecutive lattice points, the entropy  increases by $S(1)$ i.e.\ the
entropy of one lattice point,  but decreases by $S^{c}_{i}$ (for
$i=2,\dots,n+1$). Thus it is natural to think of $S^{c}_{i}$ as a
measure of the degree of correlation of a chain of lattice points of
length $i$ over and above the correlations involving subchains of length
$j$ where $j<i$.  Thus as we mentioned in the introduction,  our
$S^{c}_{n}$ is related to the index of correlation $I_{n}$  in somewhat
the same way that truncated correlation functions (sometimes known as
connected correlation functions) are related to full correlation
functions in quantum field theory and statistical mechanics
\cite{ha96,p88}.

We now use this formalism to prove Theorem~\ref{Th:entropy}  for the
case of the lattice $\mathbb{Z}$. This can then be proven to extend to
$\mathbb{Z}^{\nu}$ and $\mathbb{R}^{\nu}$ in a similar way to that in
which we proved Cases~\ref{case2},~\ref{case3} and~\ref{case4} from 
Case~\ref{case1} in Section~\ref{S:thm1}. 

Proving Theorem~\ref{Th:entropy} in the case of $\mathbb{Z}$ is
equivalent to proving that
     \begin{equation} \label{Eq:diff}
          0\leqslant S(N)- S(N-1) \text{\quad for } N\geqslant 2
     \end{equation}     
To do this, we first note that all the terms in the sum
in~\eqref{Eq:entropysum} are positive. Thus for any $n>N$, removing the
last $n-N$ terms gives us the inequality
     \begin{equation*}
         S(n)\leqslant nS(1)-\sum^{N}_{m=2}(n+1-m)S^{c}_{m}
     \end{equation*}
from which we have
     \begin{equation*}
0\leqslant \frac{S(n)}{n}\leqslant
S(1)-\frac{1}{n}\sum_{m=2}^{N}(n+1-m)S^{c}_{m}.
     \end{equation*}    
Taking the limit $n\rightarrow\infty$, we deduce that
     \begin{equation*}
         0\leqslant S(1)-\sum^{N}_{m=2}S^{c}_{m}.
     \end{equation*}
Substituting the expression for $S_{m}^{c}$ given in
Equation~\eqref{Eq:corr.entropy} into the right hand side of this
inequality, one finds that all but two of the $3(N-2)+3$ terms cancel
and one is left with~\eqref{Eq:diff}.  

We remark that actually  the above proof clearly proves a stronger
statement than our theorem, namely that  $S(N)-S(N-1)$ is greater than
or equal to the limiting mean entropy!

We also remark that it is essential for Theorem~\ref{Th:entropy} that the
full system be infinite.  For example, if instead of the one dimensional
system $\mathbb{Z}$ one were to take a closed lattice unit-circle
consisting of $n$ lattice points, as discussed in
Section~\ref{S:hexagon},  then it is obviously easy to have states
(`pure total states')  for which $S(n)=0$ while $S(m)>0$ for some $m<n$.
An amusing example of this is provided by the case where each point
around our circle corresponds to a quantum system with Hilbert space
$\H=\mathbb{C}^2$ and the pure total state is the  generalized
GHZ~\cite{ghsz90} state on the $n$-fold tensor product of $\H$ with
itself
  \begin{equation*}
     \Psi=\frac{1}{\sqrt{2}}
                |\dots\uparrow\uparrow\uparrow\dots\rangle
         +\frac{1}{\sqrt{2}}
                |\dots\downarrow\downarrow\downarrow\dots\rangle
  \end{equation*}
where $\mid\uparrow\rangle$ and $\mid\downarrow\rangle$ are a choice of
orthonormal basis for $\H$. Clearly, in this case, we would have
$S(m)=\log 2$ whenever $m<n$, but $S(n)=0$ !   

Note that if we were to attempt to consider an analogue of this example
in the case of an infinite row of lattice points, then there would be no
such difficulty because we never actually \emph{assign} an entropy to an
infinite row of lattice points. (Note though that, at least if we take
the view that all observables are local observables, it would still be
correct to assign an entropy of $\log 2$ even to the state which
formally corresponds to the above generalized GHZ state in the case
``$n=\infty$'' notwithstanding the fact that  this ``looks like'' a
vector state on an infinite tensor product of $\mathbb{C}^2$.)


\section{EPILOGUE}

One immediate consequence of Axioms (A) and (C) is that, if two regions
each have zero entropy, then both their intersection and their union
must also have zero entropy.  This might be expressed by saying: ``If
a state is pure on each of two regions, it must be pure on both their
union and intersection.''

Amongst other things, this remark further illuminates one of the
heuristic remarks (concerning Theorem 6.4 of~\cite{kw91}) made in a
paper~\cite{kw91} by Kay and Wald on quantum field theory in curved
spacetime.  (See pages 55, 99 and 105 of~\cite{kw91}.)  Namely, that it
is impossible for a state to be pure on each of two `double-wedge
regions'~\cite{kw91} but mixed on their intersection.  In fact, one of
the motivations for the present research was a desire to elucidate that
remark.

With an extension of the reasoning behind the above remark, another
result  which one can easily derive, now from our full set of axioms
(A), (B), (C) and (D), is:

\begin{thm}
  In both lattice and continuum cases, and for arbitrary dimension
  $\nu$, if the entropy of any box is zero the entropy of all boxes is zero.
\end{thm}

One may prove this either as an immediate consequence of
Theorems~\ref{Th:meanent} and~\ref{Th:entropy}, or as an easy direct
consequence of the axioms.  


\section*{ACKNOWLEDGEMENTS}
  
We thank Tony Sudbery for pointing out to us the GHZ example mentioned
at the end of Section~\ref{S:correlate}. A.~R.~Kay thanks the EPSRC
for a research studentship. 



\begin{thebibliography}{99}
    \bibitem{rr67}  D. W. Robinson and D. Ruelle,
                    ``Mean entropy of states in classical statistical
                    mechanics,''  
                    Commun. Math. Phys. \textbf{5}, 288-300 (1967).
    \bibitem{lr68}  O. E. Lanford and D. W. Robinson, 
                    ``Mean entropy of states in quantum-statistical
                    mechanics,''          
                    J. Math. Phys. \textbf{9}, 1120-1125 (1968).                
    \bibitem{al70}  H. Araki and E. L. Lieb, 
                    ``Entropy inequalities,''
                    Commun. Math. Phys. \textbf{18}, 160-170 (1970).
    \bibitem{lru73} E. L. Lieb and M. B. Ruskai, 
                    ``Proof of the strong subadditivity of
                    quantum-mechanical entropy,''
                    J. Math. Phys. \textbf{14}, 1938-1941 (1973).               
    \bibitem{l75}   E. L. Lieb,
                    ``Some convexity and subadditivity properties of
                    entropy,''  
                    Bull. Am. Math. Soc. \textbf{81}, 1-13 (1975).
    \bibitem{r94}   M. B. Ruskai,
                    ``Beyond strong subadditivity? Improved bounds on
                    the contraction of generalized relative entropy,'' 
                    Rev. Math. Phys. \textbf{6}, 1147-1161 (1994).
    \bibitem{r69}   D. Ruelle, \textit{Statistical Mechanics}
                    (Benjamin, New York, 1969).                
    \bibitem{w78}   A. Wehrl, 
                    ``General properties of entropy,''
                    Rev. Mod. Phys. \textbf{50}, 221-260 (1978).
    \bibitem{h94}   R. Horodecki, 
                    ``Informationally coherent quantum states,''
                    Phys. Lett.~A \textbf{187}, 145-150 (1994).
    \bibitem{ha96}  R. Haag, \textit{Local Quantum Physics}, 2nd ed.  
                    (Springer, Berlin, 1996).
    \bibitem{p88}   G. Parisi, \textit{Statistical Field Theory} 
                    (Addison-Wesley, Reading, MA, 1988).  
    \bibitem{ps72}  G. P\'{o}lya and G. Szeg\"{o}, \textit{Problems and
                    Theorems in Analysis I} 
                    (Springer, Berlin, 1972).                 
    \bibitem{ghsz90}   D. M. Greenberger, M. A. Horne, A. Shimony and 
                       A. Zeilinger, 
                       ``Bell's theorem without inequalities,''
                       Am. J. Phys. \textbf{58}, 1131-1143 (1990).
    \bibitem{kw91}   B. S. Kay and R. M. Wald, ``Theorems on the
                     uniqueness and thermal properties of stationary, 
                     nonsingular, quasifree states on spacetimes with 
                     a bifurcate Killing horizon,'' Physics Reports
                     \textbf{207}, 49-136 (1991).
\end{thebibliography}
\end{document}